\documentclass[prb,twocolumn,showpacs,floatfix]{revtex4}
\usepackage{graphicx}
\usepackage{amsmath}
\usepackage{amssymb}
\begin{document}

\title{Spinless Hartree-Fock model of persistent current
in rings with single scatterer: Comparison with correlated models}

\author{Radoslav \surname{N\'{e}meth}}
\affiliation{Institute of Electrical Engineering, Slovak Academy of
Sciences, D\'{u}bravsk\'{a} cesta 9, 841 04 Bratislava, Slovakia}

\author{Martin \surname{Mo\v{s}ko}}
\email{martin.mosko@savba.sk}
\affiliation{Institute of Electrical Engineering, Slovak Academy of
Sciences, D\'{u}bravsk\'{a} cesta 9, 841 04 Bratislava, Slovakia}

\date{\today}

\begin{abstract}
Using the self-consistent Hartree-Fock approximation for spinless
electrons at zero temperature, we study the persistent current of the
interacting electron gas in a one-dimensional continuous ring containing a single $\delta$ barrier.
We calculate the
persistent current
as a function of the ring circumference, magnetic flux threading the ring,
barrier strength, etc.
We compare our results with the results of correlated models like
the Luttinger liquid model and the Hubbard model
solved by means of the renormalization group.
A good agreement is found. First, the persistent current
decays with the increasing ring circumference ($L$)
faster than $L^{-1}$ and eventually like $L^{-\alpha-1}$,
where $\alpha>0$ depends only on the electron-electron interaction.
Second, the persistent current is a sine-shaped function of
magnetic flux. This sine-like dependence and in particular the universal
power law $L^{-\alpha-1}$ have sofar been believed to arise only in the
correlated many-body models. Observation of these features within the Hartree-Fock model
is a surprising result.
\end{abstract}

\pacs{73.23.-b, 73.61.Ey}
%\keywords{}

\maketitle

%%%%%%%%%%%%%%%%%%%%%%%%%%%%%%%%%%%%%%%%%%%%%%%
% Section
%%%%%%%%%%%%%%%%%%%%%%%%%%%%%%%%%%%%%%%%%%%%%%%

\section{Introduction}

Electron gas in a quantum wire is a realistic one-dimensional (1D)
electron system. By tying the wire ends to each other one creates
a 1D ring. If the ring
circumference is shorter than the electron coherence length,
one speaks about the mesoscopic ring.
An external magnetic flux applied through the opening of such
ring gives rise to an equilibrium current, known as persistent
current \cite{Imry-book}.

Observations of persistent currents in metallic
and semiconducting rings \cite{experiments} stimulated many theoretical papers focused
on the electron-electron (e-e) interaction and disorder
in such systems \cite{commentreview}. A quantitative understanding of the observed amplitude of the currents
is so far not satisfactory and the role of the e-e interaction and
disorder in mesoscopic rings need further research \cite{Imry-book,Imry-RevModPhys}.
In this work we focus on the persistent current
of interacting spinless electrons in a 1D ring with a single scatterer.
Let us review recent results for this simplified problem.

If the e-e interaction is ignored, the leading behavior
of the persistent current ($I$) as a function of the magnetic flux ($\phi$)
and ring circumference ($L$) can be derived analytically for an arbitrary
scatterer \cite{Gogolin}. It is a periodic function of $\phi$ with period $\phi_0 = h/e$,
it decays as $1/L$, and its shape is determined by
the absolute value of the transmission amplitude $t_{k_F}$ of the scatterer
at the Fermi wave vector $k_F$. In the limit of small $t_{k_F}$
it reads $I \propto L^{-1} |t_{k_F}| \sin(2\pi \phi/\phi_0)$.
These results were derived for a continuous ring \cite{Gogolin} and hold also
for the ring-shaped tight-binding lattice at half filling \cite{Meden}.

For a repulsive e-e interaction in the continuous ring,
the spinless persistent current can still be
derived analytically in the Luttinger liquid model \cite{Gogolin}. For $L \rightarrow \infty$
and for an arbitrary strength of the scatterer, the current is
$I \propto L^{-\alpha-1} \sin(2\pi \phi/\phi_0)$, where the power $\alpha>0$
is universal, depending only on the e-e interaction.

The authors of Ref. \onlinecite{Meden} obtained the spinless persistent current microscopically
by solving the Hubbard chain with nearest-neighbor hopping and interaction.
Using the renormalization group, they indeed found for a 1D
ring with a single scatterer the current which decays faster
than $1/L$ and is a sine-like function of $\phi$. The shape $I \propto \sin(2\pi \phi/\phi_0)$
and the universal scaling $I \propto L^{-\alpha-1}$
were confirmed for long chains and/or strong scatterers.

They also concluded that the sine-like shape
and the $L^{-\alpha-1}$ scaling can only be obtained in the correlated
many-body model, not in the Hartree-Fock approximation ignoring
the correlation effects. They used
a non-self-consistent Hartree-Fock approximation
and, indeed, found the qualitatively wrong results. The self-consistent
Hartree-Fock approximation, according to these authors, drives
the model with a single scatterer
into the charge-density-wave groundstate with a finite single-particle gap,
which is wrong because a single scatterer cannot change the bulk properties
of the system. Hence, no attempt has been done to
test the dependence $I \propto L^{-\alpha-1} \sin(2\pi \phi/\phi_0)$
in the self-consistent Hartree-Fock approximation. However, there is a motivation to do so.

Indeed, recently \cite{Mosko} the self-consistent Hartree-Fock
approximation has been used to study tunneling of
the weakly-interacting electron gas through a single scatterer
in a continuous 1D wire with leads.
It has been found
that there is no charge-density-wave groundstate in such model, moreover,
the tunneling probability has been found to decay with the wire length as $L^{-2\alpha}$,
in accord with correlated models \cite{Kane-92,Matveev-93}. All this suggests that an attempt to
obtain the persistent current $I \propto L^{-\alpha-1} \sin(2\pi \phi/\phi_0)$
in the self-consistent Hartree-Fock approximation might be meaningful.

We present such attempt in our present work. Using the self-consistent Hartree-Fock
approximation at zero temperature, we study the persistent current of the
interacting spinless electron gas in a continuous 1D ring with a single $\delta$ barrier.
We compare our results with the results of correlated models, with
the Luttinger liquid model \cite{Gogolin} and the microscopic Hubbard model \cite{Meden}.
A good agreement is found. First, the persistent current
decays
faster than $L^{-1}$ and eventually like $L^{-\alpha-1}$,
where $\alpha>0$ is universal.
Second, the persistent current is a sine-like function of
magnetic flux.

The paper is organized as follows. In Sect. II we discuss the Hartree-Fock model of the 1D ring.
In Sect. III we present the results for the Hartree-Fock potentials.
In Sect. IV the Hartree-Fock results for persistent currents are presented and compared
with the results of correlated models. A summary is given in Sect. V.

\section{Hartree-Fock model of the 1D ring}

We consider $N$ interacting 1D electrons with free motion along a
circular ring threaded by magnetic flux $\phi=BS=AL$, where $S$ is the
area of the ring, $B$ is the magnetic field (constant and perpendicular
to the ring area), and $A$ is the magnitude of the resulting vector
potential (circulating along the ring circumference). Such electron
system is described by Hamiltonian

\begin{eqnarray}\label{Eqs-Hamilt-1}
 \nonumber
 H &=& \sum\limits _j^N\left[\frac{\hbar^2}{2m}\left(
    -i\frac{\partial}{\partial x_j}
    +\frac{2\pi}{L}\frac{\phi}{\phi_0}\right)^2
    +\gamma\delta(x_j)\right] \\
   &+&\frac{1}{2}\sum \limits_{j\neq i}^N V(x_j-x_i)\,,
\end{eqnarray}
where $m$ is the electron effective mass, $x_j$ is the coordinate of the
$j$-th electron along the ring, $V(x_j-x_i)$ is the e-e interaction,
and $\gamma\delta(x)$ is the potential barrier due to the
scatterer. The eigenfunction
$\Psi(x_1,x_2,\dots,x_N)$ and eigenenergy $E$ obey
the Sch\"{o}dinger equation

\begin{equation}
  H\Psi=E\Psi
\end{equation}
with the cyclic boundary condition

\begin{equation}
  \Psi(x_1,\dots,x_i+L,\dots,x_N)=\Psi(x_1,\dots,x_i,\dots,x_N)
\end{equation}
for $i=1,2,\dots,N$. In the groundstate with the eigenenergy $E_0$
the persistent current reads

\begin{equation}\label{Eqs-Vseob-Prud}
  I= -\frac{\partial}{\partial \phi}E_0(\phi)\,.
\end{equation}

In the Hartree-Fock approximation, the wave function $\Psi$ is
approximated by the Slater determinant

\begin{equation}
  \Psi(x_1,\dots,x_N)=
  \frac{1}{\sqrt{N!}}\left|\begin{array}{ccc}
                          \psi_{k_1}(x_1)&\cdots&\psi_{k_1}(x_N)\\
                \vdots& \ddots&\vdots\\
                \psi_{k_N}(x_1)&\cdots&\psi_{k_N}(x_N)
                   \end{array}\right|\,,
\end{equation}
where $k_i$ is the quantum number of the $i$-th
electron. The wave functions $\psi_k(x)$ obey the
Hartree-Fock equation
\begin{multline}\label{Eqs-Schroding-1}
  \left[\frac{\hbar^2}{2m}\left(-i\frac{\partial}{\partial x}+
    \frac{2\pi}{L}\frac{\phi}{\phi_0}\right)^2+\gamma\delta(x)\right.\\
    \left.\phantom{\frac{\hbar^2}{2m}}+U_H(x)+U_F(k,x)\right]\psi_k(x)=\varepsilon_k\psi_k(x)
\end{multline}
with the boundary condition

\begin{equation}
  \psi_k(x+L)=\psi_k(x)\,,
\end{equation}
where

\begin{equation}\label{Eqs-Hartree}
  U_H(x)=\sum_{k'} \int dx' V(x-x')|\psi_{k'}(x')|^2
\end{equation}
is the Hartree potential and

\begin{multline}\label{Eqs-Fock}
  U_F(k,x)=\\
    -\frac{1}{\psi_k(x)}\sum \limits_{k'}
        \int dx' V(x-x')\psi_k(x')\psi_{k'}^*(x')\psi_{k'}(x)
\end{multline}
is the Fock nonlocal exchange term (expressed as an effective potencial
for further convenience). In the equations \eqref{Eqs-Hartree} and
\eqref{Eqs-Fock} we sum over all occupied states $k'$.

We introduce the wave functions $\varphi_k(x)$ by substitution

\begin{equation}\label{Eqs-Substit}
  \psi_k(x)=\varphi_k(x)\exp{\left(-i\frac{2\pi}{L}\frac{\phi}{\phi_0}x\right)}\,.
\end{equation}
Equations
\eqref{Eqs-Schroding-1}-\eqref{Eqs-Fock} then give the
Hartree-Fock equation

\begin{multline}\label{Eqs-Schroding-2}
  \left[-\frac{\hbar^2}{2m}\frac{d^2}{dx^2}+\gamma\delta(x)+U_H(x)+U_F(k,x)\right]\varphi_k(x)\\
  =\varepsilon_k\varphi_k(x)
\end{multline}
with the boundary condition

\begin{equation}\label{Eqs-Boundary}
  \varphi_k(x+L)=\exp\left(i2\pi\frac{\phi}{\phi_0}\right)\varphi_k(x)\,,
\end{equation}
where the potentials $U_H(x)$ and $U_F(k,x)$ are still given by
equations \eqref{Eqs-Hartree} and \eqref{Eqs-Fock}, but with $\psi$
replaced by $\varphi$.

From
equations \eqref{Eqs-Hamilt-1}-\eqref{Eqs-Boundary} the groundstate energy $E_0=\langle\Psi|H|\Psi\rangle$
can be expressed after simple algebra as

\begin{equation}
  E_0=\sum \limits _k
  \left[\varepsilon_k-\frac{1}{2}\left\langle\varphi_k\left|U_H(x)+U_F(k,x)\right|\varphi_k\right\rangle\right]\,.
\end{equation}
Setting this expression into \eqref{Eqs-Vseob-Prud} we obtain the
persistent current in the Hartree-Fock approximation. Once $E_0(\phi)$
is known, $dE_0(\phi)/d\phi$ can be calculated numerically.

As in Refs. \onlinecite{Cohen-98} and \onlinecite{Cohen-97}, we
simplify the equation \eqref{Eqs-Fock} as
\begin{equation}\label{Eqs-Cohen}
  U_F(x) \simeq - \sum \limits _{k'} \int
  dx'V(x-x'){\rm Re}\left[\psi_{k'}^*(x')\psi_{k'}(x)\right]
\end{equation}
by noticing that
$\sum_{k'} \psi_{k'}^*(x')\psi_{k'}(x) \simeq \delta(x-x')$.
This 'almost closure relation' was
tested in Fig. 2 of Ref. \onlinecite{Cohen-97},
so we do not repeat the same test here.
Our comments on the reliability of approximation (\ref{Eqs-Cohen})
are given at the end of Sect. IV.
Unlike the exact form
\eqref{Eqs-Fock}, the Fock potential \eqref{Eqs-Cohen} is local and
independent on $k$. This saves computer time and memory and
allows to study long rings.

In absence of the $\delta$ barrier, the solution of equation \eqref{Eqs-Schroding-1}
is $\psi_k(x) = e^{ikx}$ and the potentials $U_H(x)$ and $U_F(x)$ become
$x$-independent constants \cite{Mosko}. These constants, $U_H$ and $U_F$, only introduce
a rigid shift of the energy scale which has no physical effect.
Therefore,
in the following text we simply consider the potentials $U_H(x)$ and $U_F(x)$ reduced by
constants $U_H$ and $U_F$, respectively.

We now discuss the solution of equations \eqref{Eqs-Schroding-2} and
\eqref{Eqs-Boundary} assuming that the potentials $U_H(x)$ and $U_F(x)$
are known. Consider first the ring region
$x \in \langle-L/2,L/2\rangle$ as a straight-line segment of an infinite 1D wire.
Inside the segment the potential is $\gamma\delta(x)+U_H(x)+U_F(x)$,
outside we keep it zero. Therefore, the wave function outside is

\begin{equation}\label{Eqs-Leftwave}
  \varphi_k(x)=ae^{ikx}+be^{-ikx}\,,\; x<-L/2\,,
\end{equation}

\begin{equation}\label{Eqs-Rightwave}
  \varphi_k(x)=ce^{ikx}+de^{-ikx}\,,\; x>L/2\,.
\end{equation}
The amplitudes $a$ and $b$ are related to $c$ and $d$ by
\begin{equation}\label{relationabcd}
  \left(\begin{array}{c}
      c \\ d
      \end{array}
  \right)=T_0
  \left(\begin{array}{c}
      a \\ b
      \end{array}
  \right)\,,
\end{equation}
where $T_0$ is the transfer matrix

\begin{equation}\label{transfermatrix}
  T_0=\left(
      \begin{array}{cc}
      \frac{1}{t_k^*} & -\frac{r_k^*}{t_k^*} \\
      -\frac{r_k}{t_k} & \frac{1}{t_k}
      \end{array}
  \right)\,,
\end{equation}
with $t_k$ and $r_k$ being the transmission and reflection amplitudes
of the electron impinging the region $\langle-L/2,L/2\rangle$ from the left.
To come back to the ring threaded by magnetic flux, we relate $\varphi_k(-L/2)$
and $\varphi_k(L/2)$ (eqs. \ref{Eqs-Leftwave}, \ref{Eqs-Rightwave}) through
the boundary condition \eqref{Eqs-Boundary}. Combining this relation with eqs. (\ref{relationabcd})
and (\ref{transfermatrix}) we obtain the equation

\begin{equation}\label{eigenvalueeq}
  T
  \left(\begin{array}{c}
      a \\ b
      \end{array}
  \right)=\exp(i2\pi\frac{\phi}{\phi_0})
  \left(\begin{array}{c}
      a \\ b
      \end{array}
  \right)\,,
\end{equation}
where
\begin{equation}\label{transfermatrixshifted}
  T=\left(
      \begin{array}{cc}
      \frac{1}{t_k^*}e^{ikL} & -\frac{r_k^*}{t_k^*}e^{ikL} \\
      -\frac{r_k}{t_k}e^{-ikL} & \frac{1}{t_k}e^{-ikL}
      \end{array}
  \right)\,.
\end{equation}
Thus $\exp(i2\pi\phi/\phi_0)$ is the eigenvalue of matrix $T$.
The product of the eigenvalues of a matrix is given by its determinant,
which is unity in this case. The second eigenvalue is thus
$\exp(-i2\pi\phi/\phi_0)$. Their sum is equal to the matrix
trace, which gives the equation for the spectrum, \cite{Gogolin}

\begin{equation}\label{Eqs-trancedent}
  \cos\left(2\pi\frac{\phi}{\phi_0}\right)={\rm
  Re}\left[\frac{\exp(-ikL)}{t_k}\right]\,.
\end{equation}
Numerical solution  of equation \eqref{Eqs-trancedent} has to be combined with
computation of the transmission amplitude $t_k$. We compute
$t_k$ and $r_k$ in the same way as described for the open wire \cite{Mosko}.
The solution of equation \eqref{Eqs-trancedent} gives us the dependence
$k(\phi)$ and finally $\varepsilon_k(\phi)=\hbar^2k^2(\phi)/2m$.

For
each $k(\phi)$ we calculate the wave function $\varphi_k(x)$ as follows.
From eqs. \eqref{eigenvalueeq} and \eqref{transfermatrixshifted} we obtain
the amplitude

\begin{equation}
  a=\left[\frac{1}{r_k}-\frac{t_k}{r_k}e^{i(2\pi\phi/\phi_0+kL)}\right]b\,
\end{equation}
and the amplitude $b$ by normalizing the wave function. Then we express from
equation \eqref{Eqs-Leftwave}
the boundary conditions $\varphi_k(-L/2)$ and $d\varphi_k(-L/2)/dx$.
We solve the equation \eqref{Eqs-Schroding-2} with these boundary conditions
by using the same finite-difference method as described for open wire \cite{Mosko}.

Once we obtain $\varphi_k(x)$, we can also perform
the self-consistent Hartree-Fock calculation. In the first iteration step
we solve equation \eqref{Eqs-Schroding-2} for the non-interacting gas.
The resulting $\varphi_k(x)$ is used to evaluate the Hartree and Fock potentials.
These potentials are used in the second iteration step to obtain new $\varphi_k(x)$
and new potentials \cite{comment1}, etc., until the energies $\varepsilon_k$
do not change any more \cite{Cohen-97}.

In this work we apply the above Hartree-Fock scheme
to the GaAs ring with electron density $n=5 \times 10^7$ m$^{-1}$, effective mass $m=0.067$~$m_0$, and e-e
interaction
\begin{equation} \label{VeeExp}
V(x - x') = V_0 \,  e^{- \left| x - x' \right|/d},
\end{equation}
where $V_0 = 34$~meV and $d = 3$ nm. We adopt the finite-ranged interaction
(\ref{VeeExp}) due to the following reasons.
The e-e interaction in electron gas is always screened,
so a reasonably chosen screened e-e interaction improves the
Hartree-Fock approximation \cite{Ashcroft}. The e-e interaction (\ref{VeeExp})
has already been used to study the many-body 1D systems \cite{Meden-2000},
it has been shown \cite{Mosko} to
mimic the screened e-e interaction in a realistic 1D system quite well.
Finally, we want to compare our study with the correlated models
which also rely on the e-e interaction of finite range.

\section{Hartree-Fock potential}

In the following we label the transmission and reflection amplitudes of the bare $\delta$ barrier as
$\tilde{t}_k$ and $\tilde{r}_k$. It holds that \cite{Davies-98}
$\tilde{t}_k = k/(k+i\zeta)$ and $\tilde{r}_k = -i\zeta/(k+i\zeta)$,
where $\zeta = \gamma m/\hbar^2$. Since $k_F$ and $m$ are fixed,
instead of using $\gamma$ we parametrize the $\delta$ barrier by its transmission coefficient
$\left| \tilde{t}_{k_F} \right|^2$.

\begin{figure}[t!]
\begin{center}
\includegraphics[clip,width=\columnwidth]{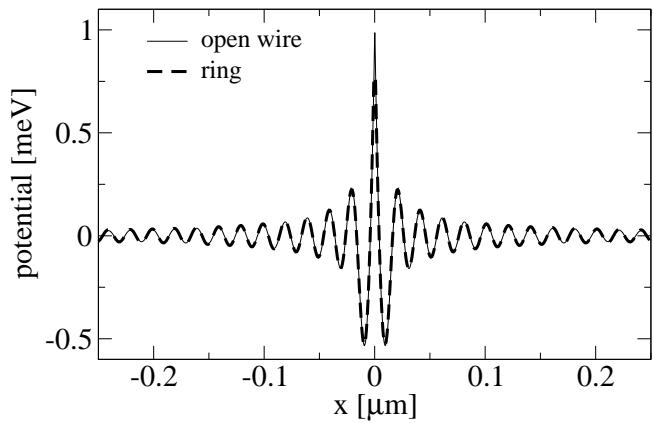}
\end{center}
\caption{
Self-consistent Hartree-Fock potential $U_H(x) + U_F(x)$ along a 1D ring, induced by
the (not shown) barrier $\gamma\delta(x)$ due to the scatterer at $x=0$.
The $\delta$ barrier is adjusted to have the bare transmission $\left| \tilde{t}_{k_F} \right|^2=0.851$
at the Fermi level ($14$ meV). The length
of the ring, $L$, is $20$ $\mu$m (in the figure
only a small region around the scatterer is shown).
The solid curve in fact involves two overlapping curves for the rings with magnetic flux
$\phi \rightarrow 0^+$ and $\phi = 0.25 \phi_0$. Notice that
we find the same curve for any $\phi$ as long as we are not far
from the $\delta$ barrier (see the next figure).
For comparison, the dashed curve shows the result of Ref. \onlinecite{Mosko},
the self-consistent Hartree-Fock potential
in a GaAs wire connected to leads,
with the $\delta$ barrier and parameters as in our ring.
} \label{Fig:Hartree-Fockpot}
\end{figure}

The solid curve in figure \ref{Fig:Hartree-Fockpot} shows
the self-consistent Hartree-Fock potential in the ring,
induced by the scatterer at $x=0$. The ring is $20$ $\mu$m long,
we show only a small region around the scatterer. The solid curve
in fact involves two curves overlapping each other, one for the ring
with very small flux $\phi \rightarrow 0^+$ and
one for the ring with $\phi = 0.25\phi_0$.
The dashed curve is the result taken from figure 2 of Ref. \onlinecite{Mosko}.
It is the self-consistent Hartree-Fock potential in the 1D GaAs wire
with the same scatterer and parameters as in our ring.
All curves exhibit the Friedel oscillations
with period $\lambda_F/2$. Since we compare the long ring and long wire
in a close vicinity of $x=0$, the agreement of the solid and dashed
curves is not surprising. As shown in the next figure, all
three curves differ
remarkably at large distances from the scatterer.

\begin{figure}[t]
\begin{center}
\includegraphics[clip,width=\columnwidth]{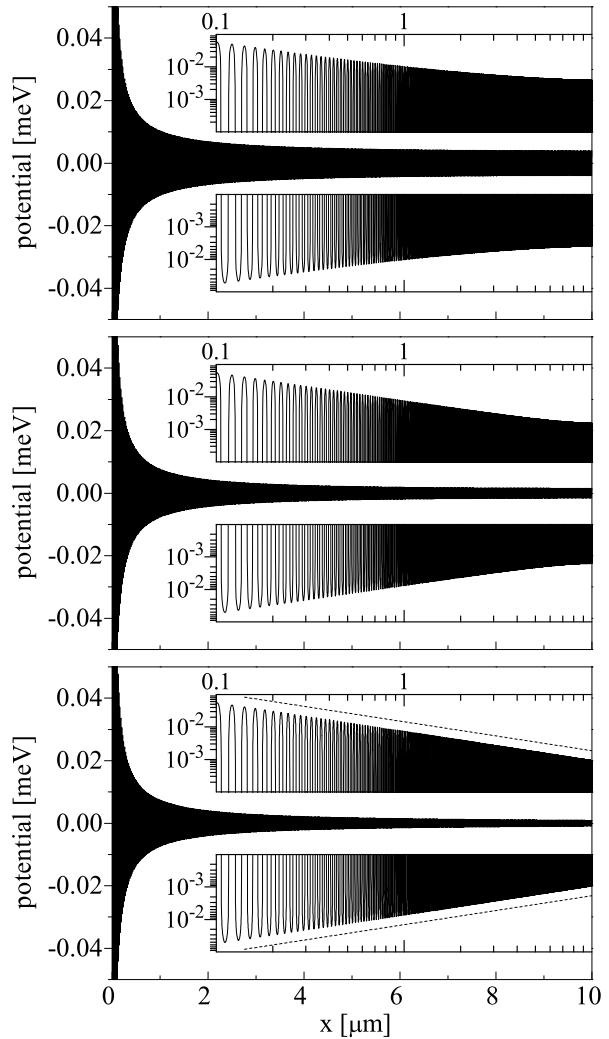}
\end{center}
\caption{The same Hartree-Fock potentials as in the preceding figure,
but along the entire half length.
The top panel and middle panel show the results
for the ring with $\phi \rightarrow 0^+$ and $\phi = 0.25\phi_0$, respectively.
For comparison, the bottom panel shows the
result for the GaAs wire with leads, taken from Ref. \onlinecite{Mosko}.
Inset to each panel shows the same data in log scale, separately for the positive (top inset)
and negative (bottom inset) data. Inset to the bottom panel shows, that the decay
of Friedel oscillations in the wire is linear in log scale. It follows the straight (dashed)
line with slope $\beta = 0.9$. A detailed analysis \cite{Mosko} shows that $\beta \rightarrow 1$
at extremely large distances.
} \label{Fig:chargedensity}
\end{figure}

Figure \ref{Fig:chargedensity} shows all three curves
along the entire half length. Here
the Friedel oscillations are too dense to be distinguishable,
but we can observe their asymptotic decay.
We see in the top and middle panel, that the amplitude of the Friedel oscillations
in the ring saturates at large distances from the scatterer.
This saturation is due to the circular shape and was
already observed in the pioneer Hartree-Fock study
of the continuous ring \cite{Cohen-98}.

However, the authors of Ref. \onlinecite{Cohen-98} discussed also the
wire with leads, but without performing any self-consistent Hartree-Fock study. They
estimated analytically for extremely strong e-e interaction, that
in the infinite wire the amplitude of the Friedel oscillations saturates
with distance similarly as in the ring. In other words,
at large distances from the scatterer
$U_H(x)+U_F(x) \propto x^{-\beta} \cos(2k_Fx+const)$, but with $\beta=0$.
This is the charge-density-wave groundstate.
In Ref. \onlinecite{Meden} this groundstate was concluded to hold for any e-e interaction strength,
but without providing explicit self-consistent Hartree-Fock results.
Nevertheless, due to this charge-density-wave groundstate,
the self-consistent Hartree-Fock method was considered to be wrong \cite{Meden}.

Recently \cite{Mosko}, the self-consistent Hartree-Fock study
has been applied to the continuous 1D wire with a single scatterer,
connected to leads. It has been shown, that the
charge-density-wave groundstate does not exist in that model at least for weak e-e interaction.
In the bottom panel of figure \ref{Fig:chargedensity} we
show the result of Ref. \onlinecite{Mosko} in order to compare with the
Hartree-Fock potential in the ring. Inset to the bottom panel
shows, that the asymptotic decay of Friedel oscillations in the wire
is linear in log scale. This implies the dependence
$\propto x^{-\beta}$ with positive $\beta>0$.
A detailed analysis \cite{Mosko} shows that $\beta \rightarrow 1$
in the limit of very long wires, i.e., there is
no tendency to a non-decaying ($\beta = 0$) periodic wave.
Saturation of the oscillation amplitude in the ring
is due to the ring geometry and does not mean any
charge-density-wave groundstate.

\section{Results for persistent current}

For noninteracting spinless electrons, the persistent current in the continuous 1D ring with a single
scatterer can be derived analytically \cite{Gogolin}. For even $N$
(considered throughout this paper) the result reads
\begin{equation} \label{I-noninteracting}
I = \frac{ev_F}{\pi L} \frac{\arccos(|\tilde{t}_{k_F}| \cos[\phi' - \pi])}
{\sqrt{1-|\tilde{t}_{k_F}|^2 \cos^2(\phi')}}|\tilde{t}_{k_F}| \sin(\phi'),
\end{equation}
where $\phi' \equiv 2\pi \phi/\phi_0$.
Equation \eqref{I-noninteracting} holds also for the tight-binding model
at half-filling \cite{Meden}. For small $\tilde{t}_{k_F}$
\begin{equation} \label{I-nonint-approx}
I = \frac{ev_F}{2 L} |\tilde{t}_{k_F}| \sin(\phi').
\end{equation}
For the repulsive e-e interaction, as already mentioned,
\begin{equation} \label{I-Luttinger}
I \propto L^{-\alpha-1} \sin(\phi')
\end{equation}
within the Luttinger-liquid model \cite{Gogolin}. Due to the
e-e interaction the current vanishes
faster than $1/L$.

These analytical predictions were tested in a numerical "experiment" of Ref. \onlinecite{Meden}.
The authors of that work studied a ring-shaped Hubbard lattice
with nearest-neighbor hopping and nearest-neighbor e-e interaction. They found for the 1D
ring with a single scatterer, that the current indeed vanishes faster
than $1/L$ and is a sine-like function of $\phi'$. Their calculation was based
on the renormalization-group (RG) technique allowing a highly reliable microscopic
treatment of correlations.

\begin{figure} [t!]
\begin{center}
\includegraphics[clip,width=8.cm]{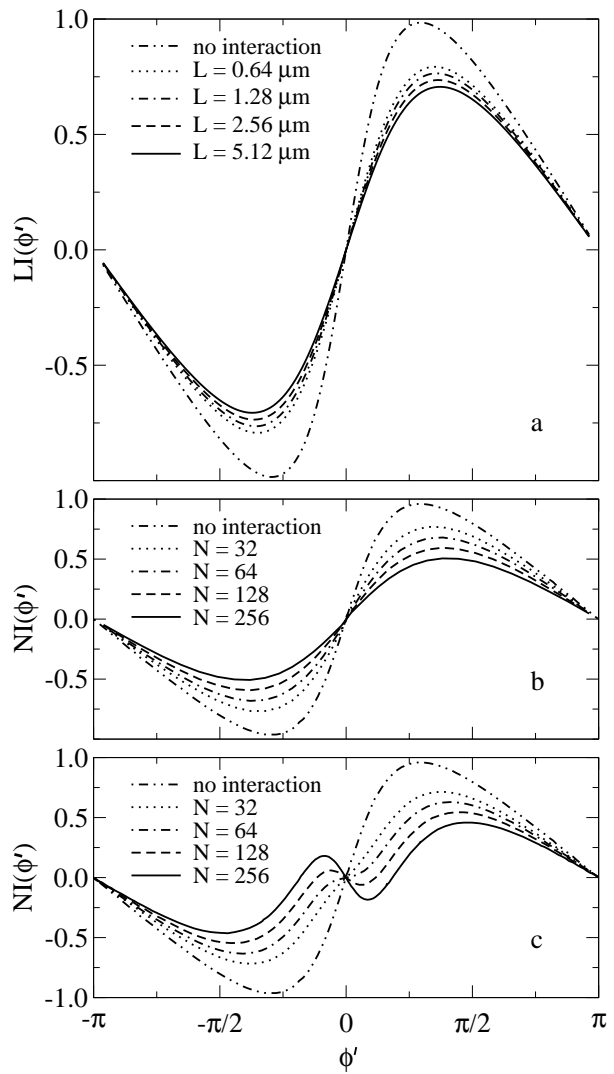}
\end{center}
\caption{Panel $a$ shows the persistent current $LI(\phi)$, calculated by our
self-consistent Hartree-Fock approach for
$\left| \tilde{t}_{k_F} \right|^2 = 0.64$ and various $L$.
For qualitative comparison, panels $b$ and $c$ show the data taken from Fig.3 of Ref. \onlinecite{Meden}.
These data were obtained \cite{Meden}
for a 1D ring-shaped Hubbard lattice
with the same $\left| \tilde{t}_{k_F} \right|^2$ and same $N$ as in panel $a$, where $N = 32, 64, \dots$
for $L = 0.64 \mu$m, $1.28 \mu$m, $\dots$, respectively.
In particular, panel $b$ shows the results of the RG solution while panel $c$ shows the results of
the (non-self-consistent) Hartree-Fock solution. The RG solution accounts for the effect of correlations.
} \label{Fig:Pot1}
\end{figure}

Let us compare the RG calculation of Ref. \onlinecite{Meden} with
our Hartree-Fock model.
In panel $a$ of figure \ref{Fig:Pot1} we show the persistent current
$LI(\phi')$ calculated by our self-consistent Hartree-Fock approach, in panel $b$ we show the
RG result of Ref. \onlinecite{Meden} in terms of $NI(\phi')$. For the purpose of comparison
we follow Ref. \onlinecite{Meden} and normalize our $LI(\phi')$ data by $ev_F/2$.
Our ring is continuous and the ring of Ref. \onlinecite{Meden} is the Hubbard lattice at half-filling.
Therefore, we cannot expect quantitative agreement.
However, we see excellent qualitative agreement. In both cases the e-e interaction
preserves the sine-like dependence on $\phi'$ and in both cases the current decays with $L$
faster than $1/L$.

\begin{figure}[t]
\begin{center}
\includegraphics[clip,width=\columnwidth]{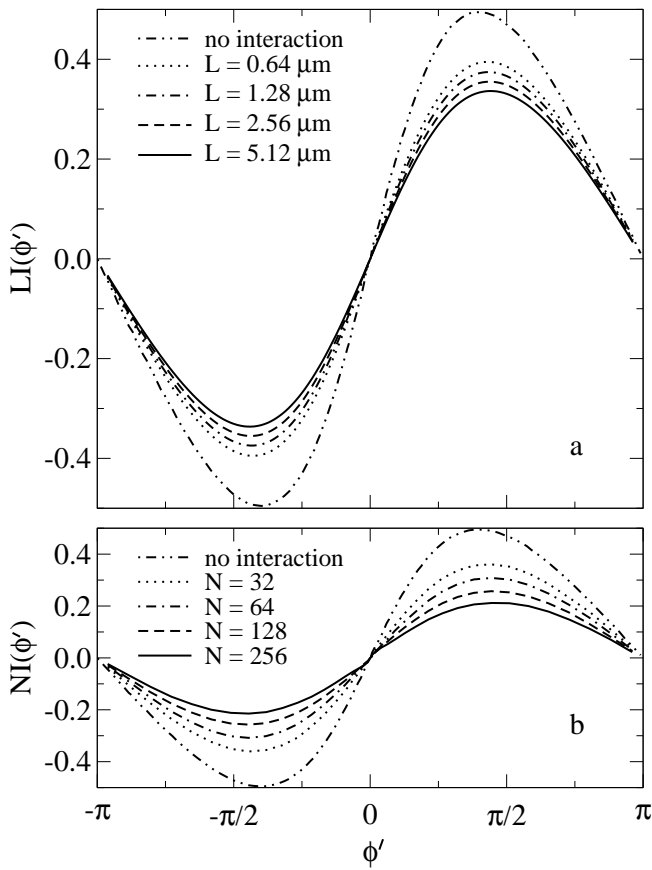}
\end{center}
\caption{Panel $a$ shows the persistent current $LI(\phi')$, calculated by our
self-consistent Hartree-Fock approach for various $L$ and
$\left| \tilde{t}_{k_F} \right|^2 = 0.22$.
For qualitative comparison, panel $b$
shows the RG data taken from Fig.4 of Ref. \onlinecite{Meden}. These data were obtained \cite{Meden}
for a 1D ring-shaped Hubbard lattice
with the same $\left| \tilde{t}_{k_F} \right|^2$ and same $N$ as in panel $a$,
where $N = 32, 64, \dots$
for $L = 0.64 \mu$m, $1.28 \mu$m, $\dots$, respectively.
} \label{Fig:Pot2}
\end{figure}

The authors of Ref. \onlinecite{Meden} also performed their own (non-self-consistent)
Hartree-Fock calculation. For the purpose of comparison,
the results of that calculation are shown in panel $c$ of Fig. \ref{Fig:Pot1}.
One can see that these results exhibit non-physical
deviations from the sine-like shape. This non-physical behavior was ascribed \cite{Meden}
to the absence of correlations in the Hartree-Fock model, but we can now ascribe it
to the absence of self-consistency (our self-consistent Hartree-Fock
results in panel $a$ clearly exhibit the sine-like shape).

In figure \ref{Fig:Pot2} we again compare our Hartree-Fock results with the RG
results of Ref. \onlinecite{Meden}, but for a stronger $\delta$ barrier. We see a similar qualitative
agreement as in Fig. \ref{Fig:Pot1}, the peak of the current is now shifted closer to
$\phi' = \pi/2$.

\begin{figure}[t]
\begin{center}
\includegraphics[clip,width=\columnwidth]{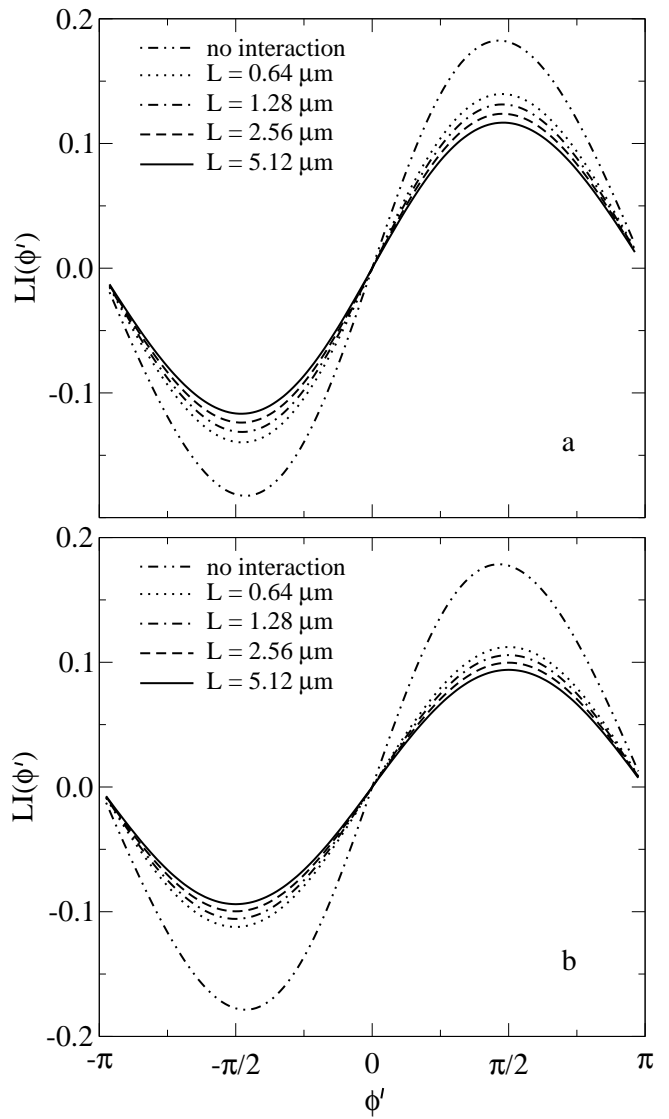}
\end{center}
\caption{Persistent current $LI(\phi')$ for
$\left| \tilde{t}_{k_F} \right|^2 = 0.03$ and various $L$. Panel $a$ shows our
self-consistent Hartree-Fock data, panel $b$ the data
typical of the Luttinger liquid (see the text).
 }
\label{Fig:Pot3}
\end{figure}

\begin{figure}[!t]
\begin{center}
\includegraphics[clip,width=\columnwidth]{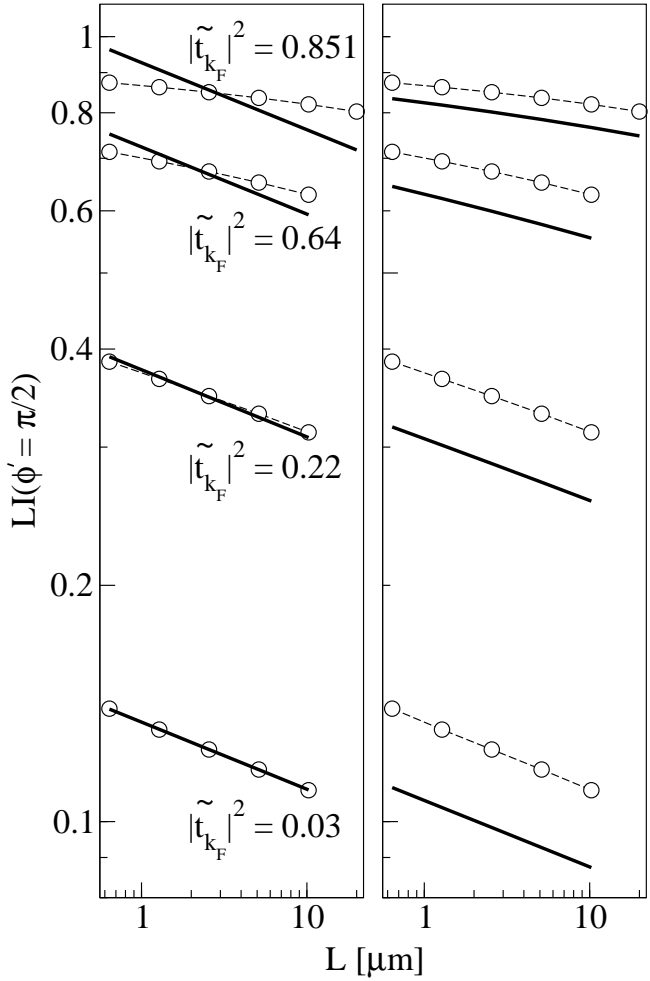}
\end{center}
\caption{Persistent current  $LI(\phi' = \pi/2)$
as a function of $L$ for various $\delta$ barriers. Our Hartree-Fock
results are shown by open symbols connected by dashed lines. The full lines in the left panel
show the scaling law (\ref{I-Luttinger-A}) with power $\alpha=0.0855$.
In this calculation the prefactor $\omega$ in eq. (\ref{I-Luttinger-A})
was adjusted to fit the Hartree-Fock data quantitatively.
The full lines in the right panel show the exact dependence (\ref{I-noninteracting})
with $\tilde{t}_{k_F}$ replaced by the exact amplitude (\ref{t-Glazman-Fermi}).
} \label{Fig:Pot4}
\end{figure}

In figure \ref{Fig:Pot3} we present the persistent current $LI(\phi')$
for the $\delta$ barrier as strong as $\left| \tilde{t}_{k_F} \right|^2 = 0.03$.
Our self-consistent Hartree-Fock results are shown in panel $a$. We want
to compare these results with the scaling law due to the Luttinger liquid model (eq. \eqref{I-Luttinger}).
To evaluate the scaling law \eqref{I-Luttinger}, we reformulate it
as follows \cite{Gogolin}. The idea is to replace the bare transmission amplitude $\tilde{t}_{k_F}$
in the non-interacting scaling law (\ref{I-nonint-approx}) by the
transmission amplitude of the interacting electron gas \cite{Matveev-93},

\begin{equation} \label{t-Glazman-Fermi}
t_{k_F} =
\frac{\tilde{t}_{k_F} \, (d/L)^{\alpha}}
     {\sqrt{ \left| \tilde{r}_{k_F} \right|^2 +
      \left| \tilde{t}_{k_F} \right|^2 \,
      (d/L)^{2\alpha} }
}\simeq \frac{\tilde{t}_{k_{F}}}{| \tilde{r}_{k_{F}}|}\left(d/L\right)^\alpha,
\end{equation}
where $d$ is the range of the e-e interaction $V(x-x')$ and
the right hand side of (\ref{t-Glazman-Fermi}) holds for small $\tilde{t}_{k_F}$ and/or large $L$.
For small $\tilde{t}_{k_F}$ one indeed obtains the scaling law
\begin{equation} \label{I-Luttinger-A}
I = \omega L^{-\alpha-1} \sin(\phi'),
\end{equation}
where $\omega = e v_F \left| \tilde{t}_{k_F} \right| d^{\alpha}/ 2 \left| \tilde{r}_{k_F} \right|$.
The formula (\ref{t-Glazman-Fermi}) was derived \cite{Matveev-93} by means of the RG method
in the limit of weak e-e interaction ($\alpha \ll 1$). In that limit
$\alpha$ is given by \cite{Matveev-93}

\begin{equation} \label{alpha-Glazman}
\alpha = \frac{V(0)-V(2k_F)} {2\pi \hbar v_F},
\end{equation}
where $V(q)$ is the
Fourier transform of the e-e interaction $V(x-x')$.
The Fourier transform of our interaction \eqref{VeeExp} reads
$V(q) = 2V_0 d/(1+q^2d^2)$. We set this formula and our parameters
into equation \eqref{alpha-Glazman}. We obtain $\alpha = 0.0855$.
We evaluate the formula (\ref{I-Luttinger-A})
and we show this result in panel $b$ of figure \ref{Fig:Pot3}.
Clearly, this result and the Hartree-Fock result of panel $a$
are in excellent qualitative and good quantitative
agreement.

In figure \ref{Fig:Pot4} we show the persistent current
$LI(\phi' = \pi/2)$
as a function of $L$ for various $\delta$ barriers. Our Hartree-Fock
results are shown by open symbols connected by dashed lines. The full lines in the left panel
show the scaling law (\ref{I-Luttinger-A}) with the power $\alpha=0.0855$ obtained from the
universal formula (\ref{alpha-Glazman})
and with the factor $\omega$ adjusted to fit the open symbols quantitatively.
We see
that the Hartree-Fock data reproduce the universal law $I \propto L^{-\alpha-1}$
in the limit of small $\tilde{t}_{k_F}$.
Of course, the
power law (\ref{I-Luttinger-A}) is asymptotic and cannot agree
with the Hartree-Fock data for any $\tilde{t}_{k_F}$. A good agreement is obtained
if we plot directly the exact formula (\ref{I-noninteracting}) with $\tilde{t}_{k_F}$
replaced by the exact amplitude (\ref{t-Glazman-Fermi}).
This exact dependence is shown
in a full line in the right panel of figure \ref{Fig:Pot4}.
The full lines and Hartree-Fock data now agree quite well for any $\tilde{t}_{k_F}$
and scale as $I \propto L^{-\alpha-1}$ for small $\tilde{t}_{k_F}$. Why a small quantitative
difference appears is explained below.

\begin{figure}[!t]
\begin{center}
\includegraphics[clip,width=\columnwidth]{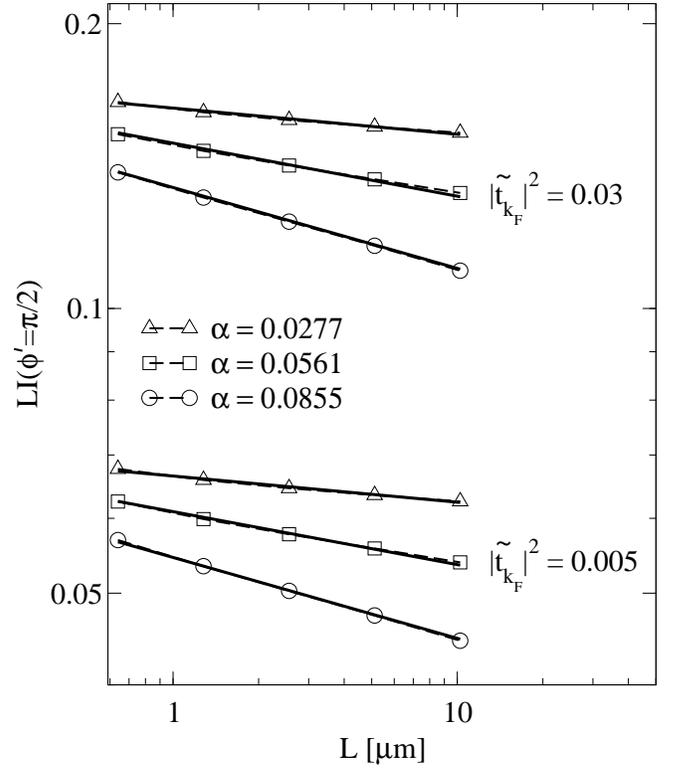}
\end{center}
\caption{Persistent current  $LI(\phi' = \pi/2)$
versus $L$ for various $\delta$ barriers and
various e-e interaction strength (various $\alpha$). Our Hartree-Fock
results are shown by open symbols connected by dashed lines. The full lines
show the scaling law (\ref{I-Luttinger-A}) with the factor $\omega$
adjusted to fit the Hartree-Fock data quantitatively.
In this calculation the powers  $\alpha = 0.0277$ and $\alpha = 0.0561$ correspond to the
the e-e interaction amplitudes $V_0 = 11$~meV and $V_0 = 22.3$~meV, respectively. Note that
we can keep the same value of $\alpha$ by using the e-e interaction (\ref{VeeExp})
with many different combinations of $V_0$ and $d$. However, as long as $\alpha$ remains the same,
the choice
of $V_0$ and $d$ has no effect on our results (Hartree-Fock potentials, eigenenergies, currents).
} \label{Fig:Pot7}
\end{figure}

Eventually, for small enough $\tilde{t}_{k_F}$ our Hartree-Fock data always follow
the scaling law $I \propto L^{-\alpha-1}$.
This is demonstrated in Fig.\ref{Fig:Pot7} for several $\tilde{t}_{k_F}$ and $\alpha$.

Now we add a few arguments on the reliability of approximation (\ref{Eqs-Cohen}).
Our Hartree-Fock results
systematically agree with the results of correlated models. This would hardly be possible in
a wrong Hartree-Fock model. In particular, we have recovered
the power law $I \propto L^{-\alpha-1}$, where $\alpha$ is given by universal expression (\ref{alpha-Glazman}).
It is highly improbable that such nontrivial physical law
could be a purely accidental consequence of wrong
approximation in the Hartree-Fock model.

Since it relies\cite{Cohen-97} on the 'almost closure relation'
$\sum_{k'} \psi_{k'}^*(x')\psi_{k'}(x) \simeq \delta(x-x')$,
approximation (\ref{Eqs-Cohen}) in fact underestimates the exact Fock potential (\ref{Eqs-Fock})
and overestimates the electron transmission and persistent current.
Using of the exact Fock potential (\ref{Eqs-Fock}) would just remove
a small quantitative difference between the Hartree-Fock and correlated results
in the right panel of Fig. \ref{Fig:Pot4}.

Why the approximation (\ref{Eqs-Cohen}) works so well can be qualitatively understood as follows.
It can be seen to be a perfect approximation of equation (\ref{Eqs-Fock})
in the extreme case $V(x - x') \propto \delta(x - x')$. Thus, the shorter the range of the
interaction $V(x - x')$ the better the approximation \eqref{Eqs-Cohen}.
Our interaction (\ref{VeeExp}) is short-ranged
and in this respect more appropriate than the bare Coulomb interaction
for which the approximation (\ref{Eqs-Cohen}) seems to work well. \cite{Cohen-98,Cohen-97}
For $V(x - x') \propto \delta(x - x')$ the Hartree and Fock potentials cancel to zero,
the exponentially decaying e-e interaction (\ref{VeeExp}) still produces a weak Hartree-Fock potential.

\section{Concluding remarks}

In conclusion, using the self-consistent Hartree-Fock
approximation at zero temperature, we have studied the persistent current of the
weakly-interacting spinless electron gas in a continuous 1D ring with a single $\delta$ barrier.
We have compared our results with the results of correlated models, with
the Luttinger liquid model \cite{Gogolin} and the microscopic Hubbard model \cite{Meden}.
We have found a good agreement. In particular, in the limit of strong $\delta$ barriers
our self-consistent Hartree-Fock
approximation reproduces the scaling of the persistent current in the form
$I \propto L^{-\alpha-1} \sin(2\pi \phi/\phi_0)$, where $\alpha$
is a universal power, depending only on the e-e interaction.
We can therefore conclude, that the universal power law is not exclusively
the consequence of correlations; it can also be obtained in the self-consistent
Hartree-Fock approximation.

We believe that the self-consistent Hartree-Fock approximation
would confirm the $L^{-\alpha-1}$ scaling also in the limit of weak $\delta$ barriers,
where the $L^{-\alpha-1}$ scaling arises only at extremely
large $L$. Unfortunately, our present computer equipment is too slow for this task.

The self-consistent Hartree-Fock
approximation has recently \cite{Mosko} been used to study tunneling of
the weakly-interacting electron gas through a single scatterer
in a continuous 1D wire with leads. It has been found \cite{Mosko} in the limit of
strong $\delta$ barriers, that the
tunneling probability decays with the wire length as $L^{-2\alpha}$,
in accord with correlated models \cite{Kane-92,Matveev-93}. A quite similar agreement with correlated
models shows our present Hartree-Fock study by reproducing
the $L^{-\alpha-1}$ scaling of persistent
current \cite{Gogolin,Meden}.

\begin{acknowledgments}
M.M. was supported by the APVT grant APVT-51-021602,
R.N. by the VEGA grant 2/3118/23.
\end{acknowledgments}

%%%%%%%%%%%%%%%%%%%%%%%%%%%%%%%%%%%%%%%%%%%%%%%%%%%%%%%%%%%%%%%%%%%
% Bibliography
%%%%%%%%%%%%%%%%%%%%%%%%%%%%%%%%%%%%%%%%%%%%%%%%%%%%%%%%%%%%%%%%%%%


\begin{thebibliography}{99}

\bibitem{Imry-book} Y.~Imry, \emph{Introduction to Mesoscopic Physics} (Oxford
    University Press, Oxford, UK, 2002).

\bibitem{experiments}
L.~P. Levy, G. Dolan, J. Dunsmuir, and H. Bouchiat, Phys. Rev. Lett. \textbf{64}, 2074 (1990);
V. Chandrasekar, R.~A. Webb, M.~J. Brady, M.~B. Ketchen, W.~J. Gallagher, and A. Kleinsasser, Phys. Rev. Lett.
\textbf{67}, 3578 (1991);
D. Mailly. C. Chapelier, and A. Benoit, Phys. Rev. Lett. \textbf{70}, 2020 (1993);
E.~M.~Q. Jariwala, P. Mohanty, M.~B. Ketchen, and R.~A. Webb, Phys. Rev. Lett. \textbf{86}, 1594 (2001);
W. Rabaud, L. Saminadayar, D. Mailly, K. Hasselbach, A. Beno\^{i}t, and B. Etienne, Phys. Rev. Lett. \textbf{86},
3124 (2001).

\bibitem{commentreview}
Reviews of this issue can be found in S. Viefers, P. Koskinen, P.~S. Deo, and M. Manninen, Physica E \textbf{21},
1 (2004), in U. Eckern, and P. Schwab, J. Low Temp. Phys. \textbf{126}, 1291 (2002), and in Ref. \onlinecite{Imry-book}.

\bibitem{Imry-RevModPhys}
Y.~Imry and R.~Landauer, Rev. Mod. Phys \textbf{71}, S306
(1999).

\bibitem{Gogolin}
A.~O. Gogolin and N.~V. Prokof'ev, Phys. Rev. B \textbf{50}, 4921 (1994).

\bibitem{Meden}
V. Meden and U. Schollw\"ock,
Phys. Rev. B \textbf{67}, 035106 (2003).

\bibitem{Mosko}
M. Mosko, A. Gendiar, P. Vagner, and Th. Schappers, submitted to Phys. Rev. B.

\bibitem{Kane-92}
C.~L. Kane and M.~P.~A. Fisher, Phys. Rev. Lett. \textbf{68}, 1220
(1992).

\bibitem{Matveev-93}
K.~A. Matveev, D. Yue, and L.~I. Glazman, Phys. Rev. Lett. \textbf{71}, 3351
(1993); D. Yue, L.~I. Glazman, and K.~A. Matveev, Phys. Rev. B \textbf{49}, 1966
(1994).

\bibitem{Cohen-98}
A. Cohen, K. Richter, and R. Berkovits, Phys. Rev. B \textbf{57}, 6223
(1998).

\bibitem{Cohen-97}
A. Cohen, R. Berkovits, and A. Heinrich, Int. J. Mod. Phys. B \textbf{11}, 1845
(1997).

\bibitem{comment1}
Inspired by the Schr\"odinger/Poisson solver for inversion Si
layers [T. Ando, A.~B. Fowler, and F. Stern, Rev. Mod. Phys.
\textbf{54}, 437 (1982)], we use a similar iterative trick.

\bibitem{Ashcroft}
N.~W. Ashcroft, and N.~D. Mermin,\emph{Solid state
physics}, Sounders College Publishing, Ed. D. G. Crane,
 Cornell University, USA 1976.

\bibitem{Meden-2000}
V. Meden, W. Metzner, U. Schollw\"ock, O. Schneider, T. Stauber, and K. Sch\"onhammer,
Europhys. J. B \textbf{16}, 631 (2000).

\bibitem{Davies-98}
J. H. Davies, \emph{The Physics of Low-Dimensional Semiconductors:
An introduction} (Cambridge University Press, Cambridge, UK, 1998).


\end{thebibliography}
\end{document}